\documentclass{kapproc} 

\usepackage{procps} 

\usepackage[dvips]{graphicx}

\upperandlowercase

\normallatexbib 

\begin{document}

\articletitle{New Physics In Clusters of Galaxy}

\articlesubtitle{A New Dark Matter Component ?}

\author{M. Douspis}
\affil{Laboratoire d'astrophysique de Toulouse Tarbes}
\email{douspis@ast.obs-mip.fr}

\begin{abstract}

The latest cosmological observables analyses seem to converge to a concordant
view of the cosmological model: namely the power law $\Lambda$-CDM.
The recent WMAP results comfort this new standard model. Nevertheless,
some degeneracy in the CMB physics do not allow one to exclude
alternative models. A combined analysis with other cosmological
observations is thus needed. An example of such work is shown here,
focusing on the abundance of local clusters. The latter is a
traditional way to derive the amplitude of matter fluctuations, but
which suffers from a lack of accurate knowledge of their masses. Here
we show that the mass temperature relation can be specified for any
cosmological model from consistency arguments, removing most of this
uncertainty. This allows to obtain an estimation of the amplitude of
matter fluctuations with an accuracy of 5\%. Quite remarkably, this
amplitude can be also tightly constrained from existing CMB
measurements. However, the amplitude inferred in this way in a
concordance model ($\Lambda$-CDM) is significantly larger than the
value derived from x-ray clusters. Such a discrepancy may reveal the
existence of a new dark component in the Universe. It may
alternatively reveal a significant depletion of the gas during the
formation of clusters. In all cases, an essential element of clusters
formation history seems to be missing.

\end{abstract}

\begin{keywords}
Cosmology -- CMB -- Clusters of galaxy -- Parameter estimation
\end{keywords}

\section*{Introduction}

The cosmological scenario seems more and more well defined.  Recent
cosmological analyses have lead to the so-called concordance
model. The latter passes a lot of cosmological tests. Nevertheless,
some observations seem not to be concordant. Furthermore, constraints
on the concordance model are obtained in some (well motivated)
particular framework. Some alternatives are thus still possible and
the conclusions may be drawn with care.
In a first section we will examine the concordance model through the
constraints and degeneracies given by the study of CMB anisotropies on
cosmological parameters. We then show how some combined analyses of
different cosmological observations could lead to a more precise view
of the concordance model. After few remarks on the dark matter
candidates, we will focus on the amplitude of fluctuations and the
need for new physics (or dark matter component) to reconcile clusters
and CMB constraints.

\section{The concordance model}

\subsection{From CMB}
\subsubsection{Constraints}

The CMB anisotropies observations are used since their discovery in
the early nineties by COBE-DMR to put constraints of cosmological
parameters. The detection of the Sachs-Wolfe plateau in the angular
power spectrum (APS hereafter) in 1992 allows one to put constraints
on the shape of the initial power spectrum (IPS hereafter) of
fluctuations by constraining its amplitude and slope \cite{gorski}.
Then the first hint for the presence of a first acoustic peak in the
APS around the degree scale (multipole moment $\ell \sim 200$) was
enough for excluding low density models (open models) and comforting
inflationary predictions for a flat Universe (no curvature)
\cite{lineweaver}. Then new observations were made from ground,
balloon and recently satellite again, with a large range of
instruments, strategies, frequencies, resolutions. All these data were
used progressively to make better and better constraints on
cosmological parameters. Today around thirty experiments have observed
the microwave sky and the last one opens a new generation of
observation. WMAP 2003 release (first release) marked a step in CMB
anisotropies studies \cite{wmap}. The APS derived from full sky maps
both in temperature and in cross-polarization helped in constructing a
better convergence to the new standard $\Lambda$-CDM model. WMAP
becomes a reference by constraining most of the cosmological parameter
by itself. Such result, shown in table 1 and Figure 1, was obtained in
the following framework: a cosmic scenario in which the anisotropies
are due to some Gaussian random and adiabatic fluctuations evolving in
a cold dark matter and dark energy dominated (topologically trivial)
Universe, in which the initial power spectrum is a power law
caracterised by its amplitude ($A$) and slope ($n$). Further
hypothesis (or priors on the parameters) assumed are summarized in
Table 1. In such (well motivated) scenario, most of the errors on the
cosmological parameters are smaller than 10\%.

\begin{figure}[ht]
\centering
\includegraphics[width=9cm]{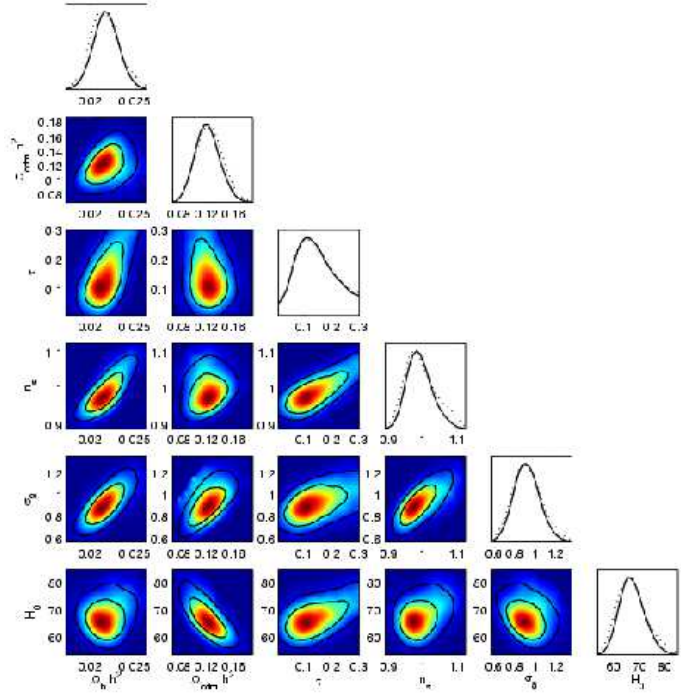}
\caption{Constraints from WMAP as 2 dimensional likelihood contours (from \cite{fosalba})}
\end{figure}

\begin{table}[ht]
\caption[Parameters from WMAP]
{Parameters from WMAP}
\begin{tabular*}{\textwidth}{@{\extracolsep{\fill}}lcc|c}
\sphline
\it Parameter&\it symbol &\it Value with errors &\it Hypothesis\cr
\sphline
Baryon density      & $\Omega_bh^2$ & $ 0.024 \pm 0.001$      &$\tau < 0.3$\cr
Dark matter density & $\Omega_bh^2$ & $0.14 \pm 0.02$         &$h > 0.5 $\cr
Hubble constant     & $h$           & $0.72 \pm 0.05$         &$\Omega_{tot}=1$\cr
Optical depth       & $\tau$        & $0.166^{0.076}_{0.071}$ &$P_0(k) \propto k^n$\cr
Spectral index      & $n$           & $0.99 \pm 0.04$         &$$\cr
\sphline
\end{tabular*}
\begin{tablenotes}
\end{tablenotes}
\end{table}
\inxx{captions,table}

\subsubsection{Degeneracies}

Even in such well defined scenario some degeneracies remain. Some
parameters have the property of changing the APS, when varying, in an
opposit manner from other (combinations of) parameters at first
order. For example, an early reionisation will suppress power at smale
scales in the temperature APS, whereas a blue power spectrum (index
$n$ bigger than 1) will enhance the power at small scale relative to
the large scales. Even at WMAP sensitivity, such effects are still
visible (elongated ellipses in Figure 1).

If some of the hypothesis made in the above scenario are relaxed more
degeneracies appear. That is the case for the one between the Hubble
constant $H_0$ and the total density if the Universe (or curvature)
$\Omega_{tot}$.  The Figure 2 shows in blue the confidence intervalles
obtained with WMAP in the $H_0, \Omega_{tot}$ plane without
assumptions on these parameters. In red are the contours when a prior
$H_0 > 50$ km/s/Mpc is assumed.  Such example shows that the CMB by
itself do not prove the flatness of the Universe. Nevertheless the CMB
data excluded drastically open low density models.

\begin{figure}[ht]

\includegraphics[width=6cm]{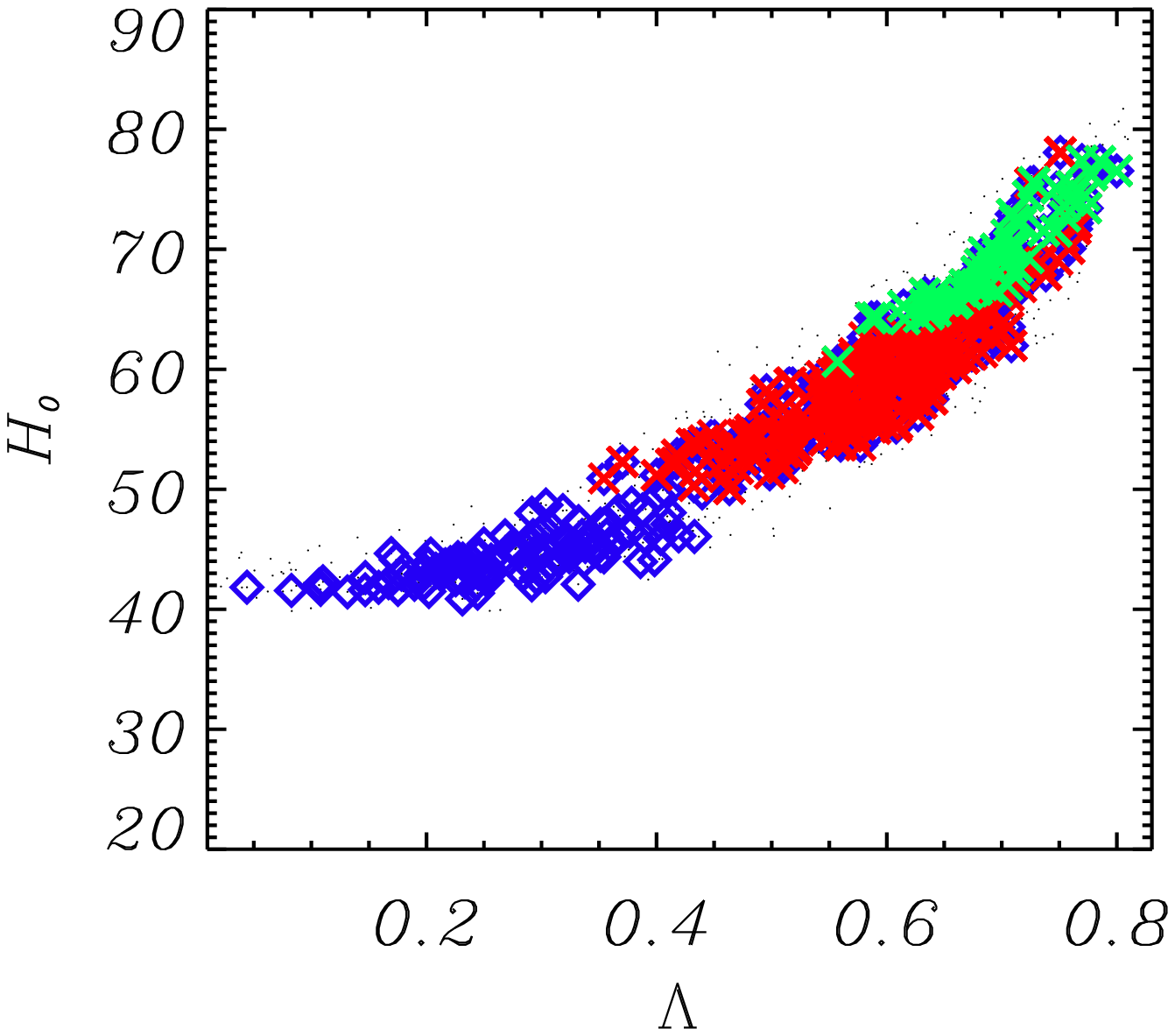}\includegraphics[width=6cm]{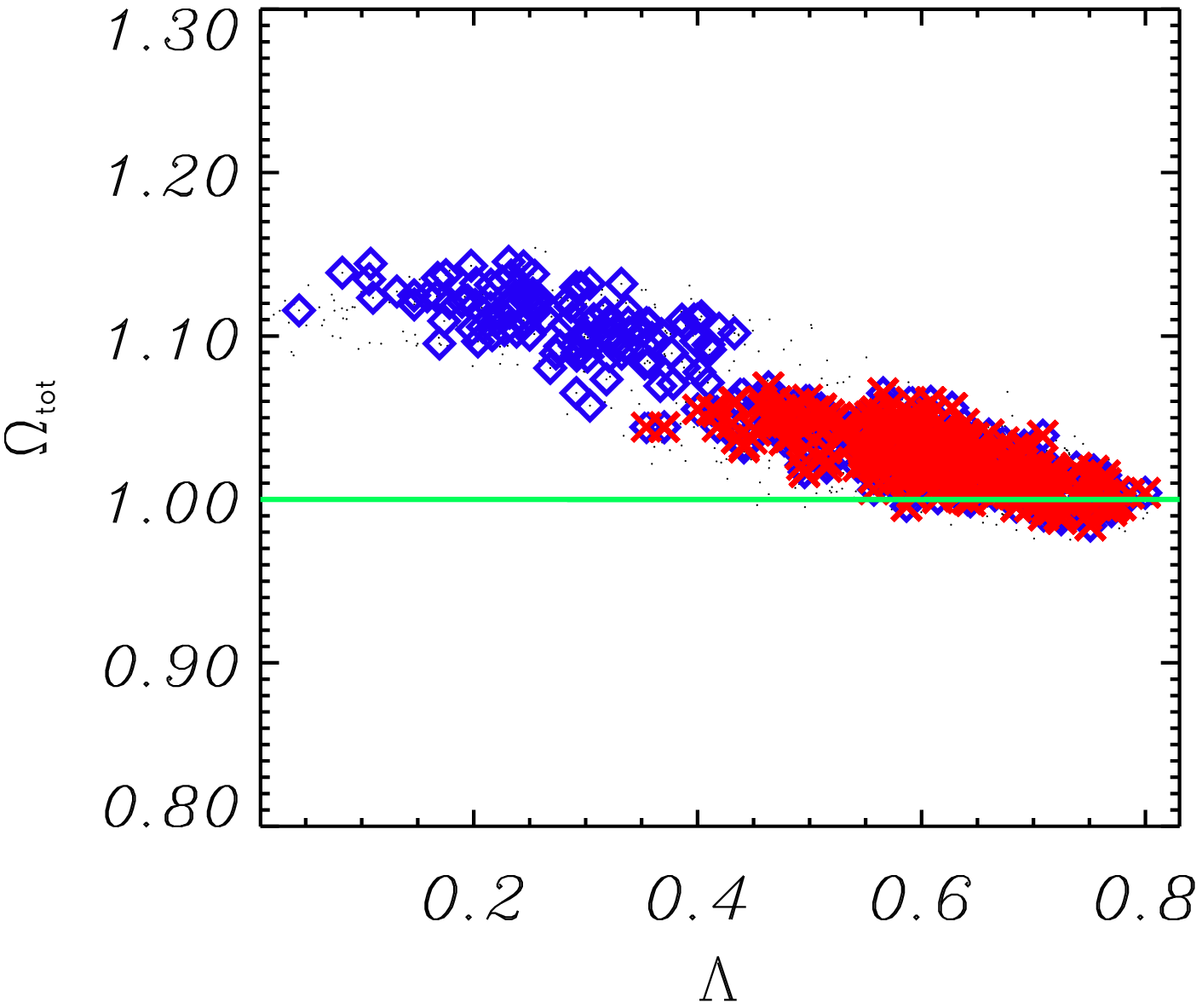}
\caption{Curvature degeneracy: constraints in the ($H_0, \Lambda$) and ($\Omega_{tot},\Lambda$) planes. In blue, analyses without priors on $H_0$ nor $\Omega_{tot}$. In red, analyses with $H_0 > 50 \rm km/s/Mpc$ prior. In green, analyses with   $\Omega_{tot}=1$ prior.}
\end{figure}

Another kind of degeneracies appear when more freedom is let to the
initial conditions. The APS is the result of the evolution in a
particular cosmology of some initial conditions such as the IPS.  The
APS could be written as follows:
\begin{equation} 
C_\ell = \int P_0(k) \Delta^2_{\ell,k} \frac{dk}{k}
\end{equation} 
where $P_0(k)$ is the IPS and $\Delta$ is the
transfer function depending on the cosmological parameters such as the
densities, the Hubble constant, etc.  As we have seen the ``standard''
scenario considers power law IPS. If such hypothesis is relaxed, then
the constraints on cosmological parameters weaken. Figure 3 show an
example from \cite{BDRS} where two APS are drawn over WMAP data. The
dotted model is the best power law $\Lambda$-CDM model obtained in
\cite{spergel}, whereas the blue solid line model in a pure matter
 dominated flat model ($\Omega_\Lambda=0$) with a IPS caracterised
 with two different slopes at large and  small scales. These two models
 are thus equiprobable as seed of the WMAP observed APS.

\begin{figure}[ht]

\includegraphics[width=6cm]{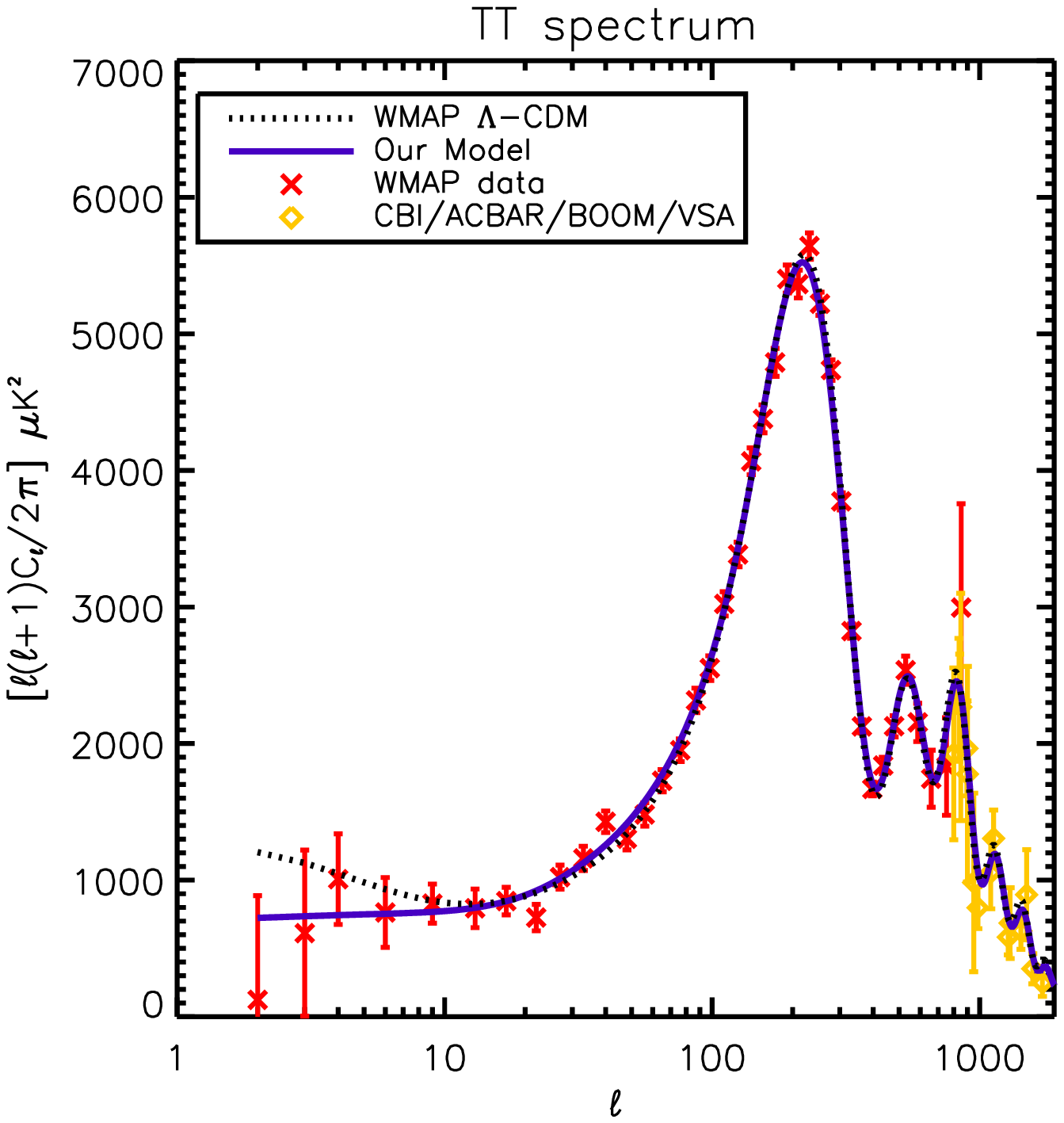}\includegraphics[width=6cm]{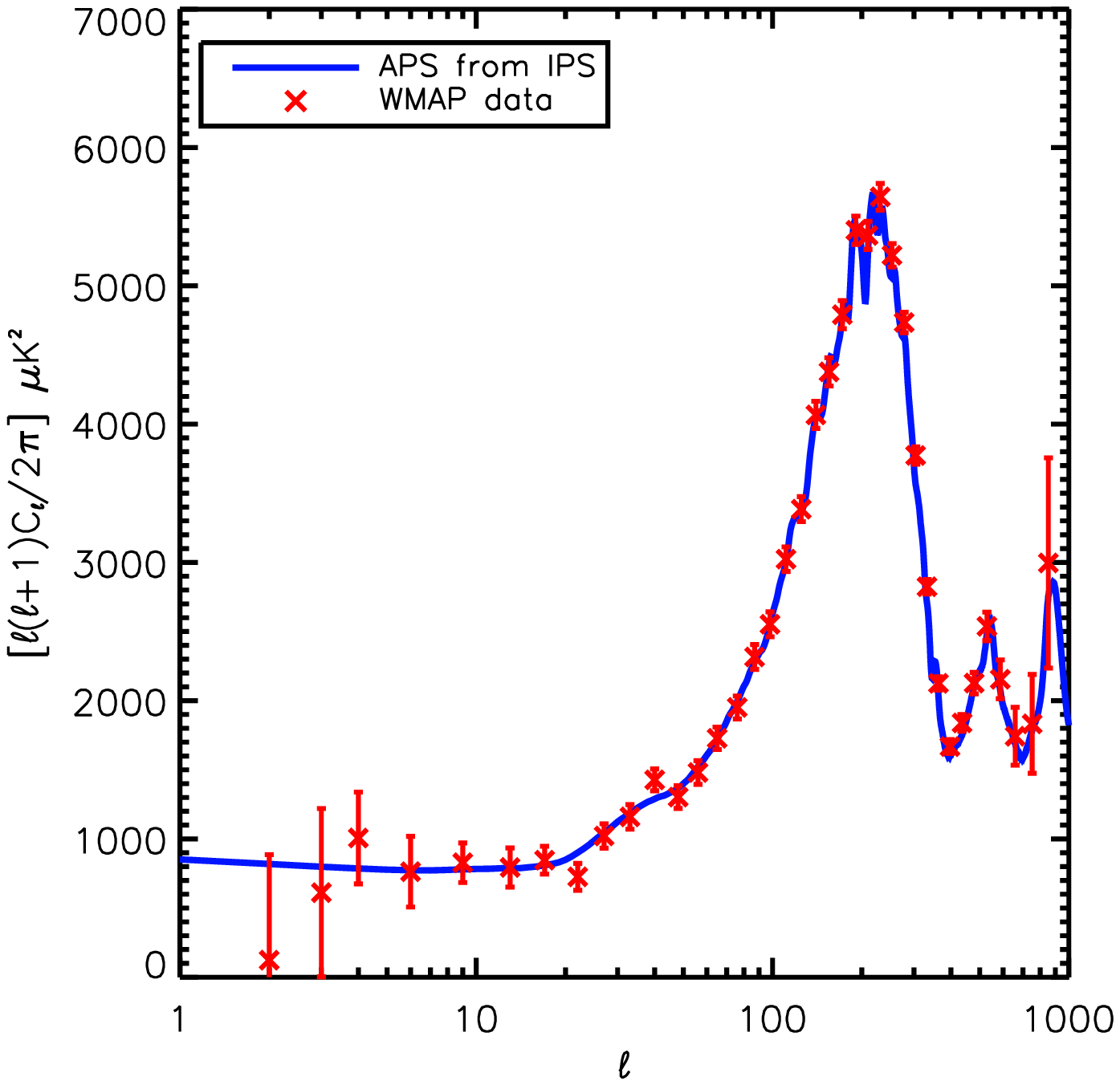}
\caption{left: Angular power spectra from power law $\Lambda$-CDM and broken power law de Sitter Universe. right: Angular power spectrum from the reconstructed initial power spectrum shown in Figure 4 (left).}
\end{figure}

As noted by the WMAP team, few outliers in the observed APS make some
strong contributions in the global $\chi^2$ which is consequently
subsequently high. These outliers, expected to be due to some
experimental effects not well substracted could also have cosmic
origin. Some features in the IPS could for example produce such
deviations from the theoretical APS. Under the assumption that these
outliers are cosmological signal, Tocchini--Valentini et
al. (\cite{TDS}) have inverted the relation in equation 1 to retrieve
the IPS from the APS (by assuming concordance cosmological
parameters). The results is plotted in Figure 4 (left). The IPS
derived from WMAP $C_\ell's$ (in blue with error bars en green)
exhibit deviations from a simple power law (black $\sim$ horizontal
line).  If such an IPS is now evolved to present time one have the
corresponding matter power spectrum that can be confronted to
observations. As shown in Figure 4 (center), the deviations in the IPS
produce deviations in the matter power spectrum that are in agreement
with the features detected by the Sloan Digital Sly Survey (SDSS,
\cite{sdss}) comforting the idea of cosmological origin of the
deviations in the WMAP APS.  The method used by Tocchini-Valentini et
al. could be applied whatever the cosmology is. Given the degeneracy
illustrated by Equation 1, one can retrieve a IPS able to fit WMAP
$C_\ell's$ for a purely matter dominated Universe. Figure 4 (right)
shows the IPS retrieved in such a cosmology ($\Omega_\Lambda=0$).

The CMB anisotropies studies can bring a lot of information on the
cosmological scenario. Strong constraints could be put if a reasonable
framework is assumed. Nevertheless, degeneracies remain in more
general approaches. The physics  of the CMB anisotropy imply
degeneracies in the shape of the APS. Moreover, the initial conditions
remain degenerated with the evolution of the Universe (see also
\cite{isoc}).

\begin{figure}[ht]

\includegraphics[width=4cm]{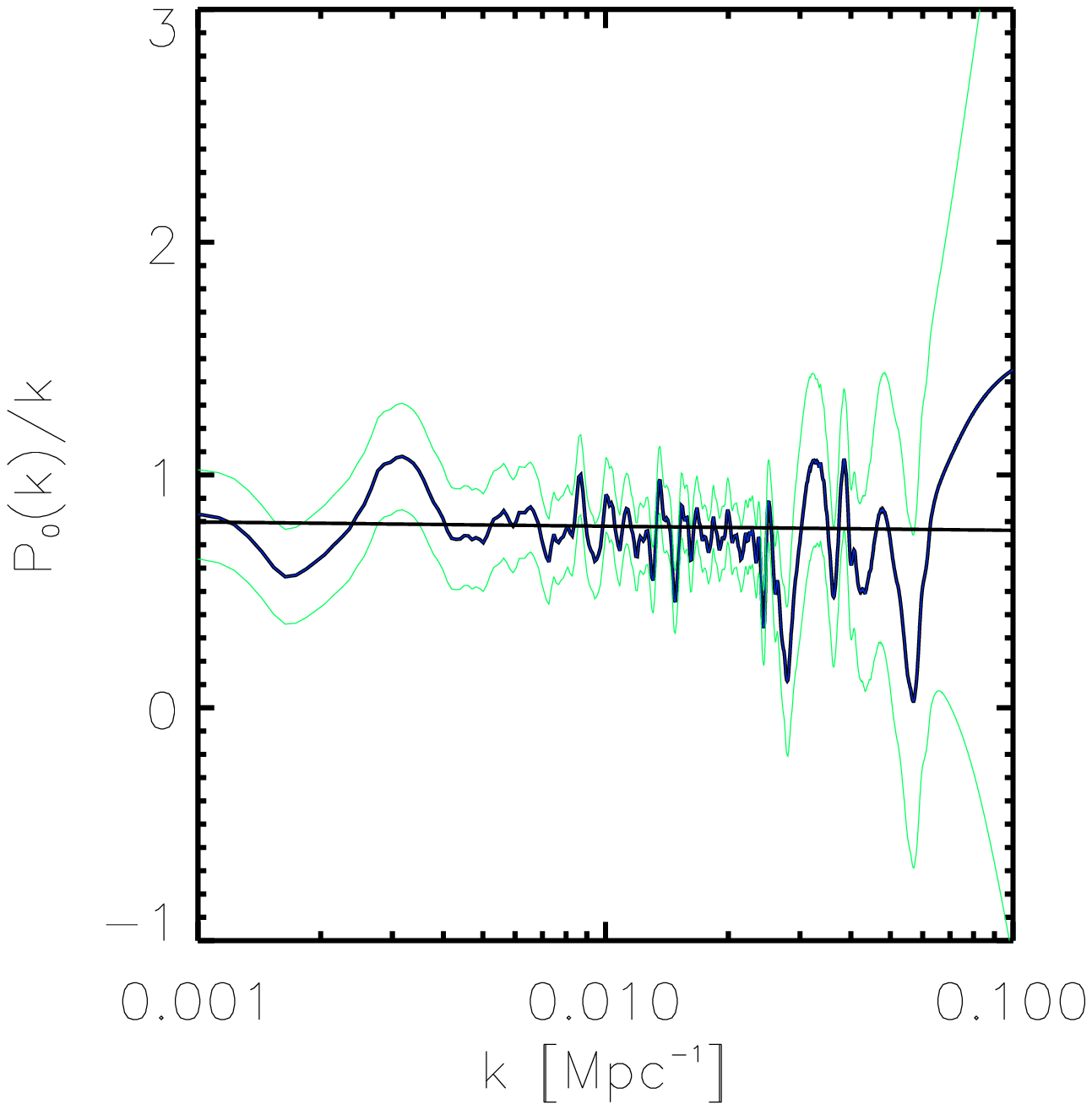}\hspace*{3.3cm}\includegraphics[width=5cm, height=3.8cm]{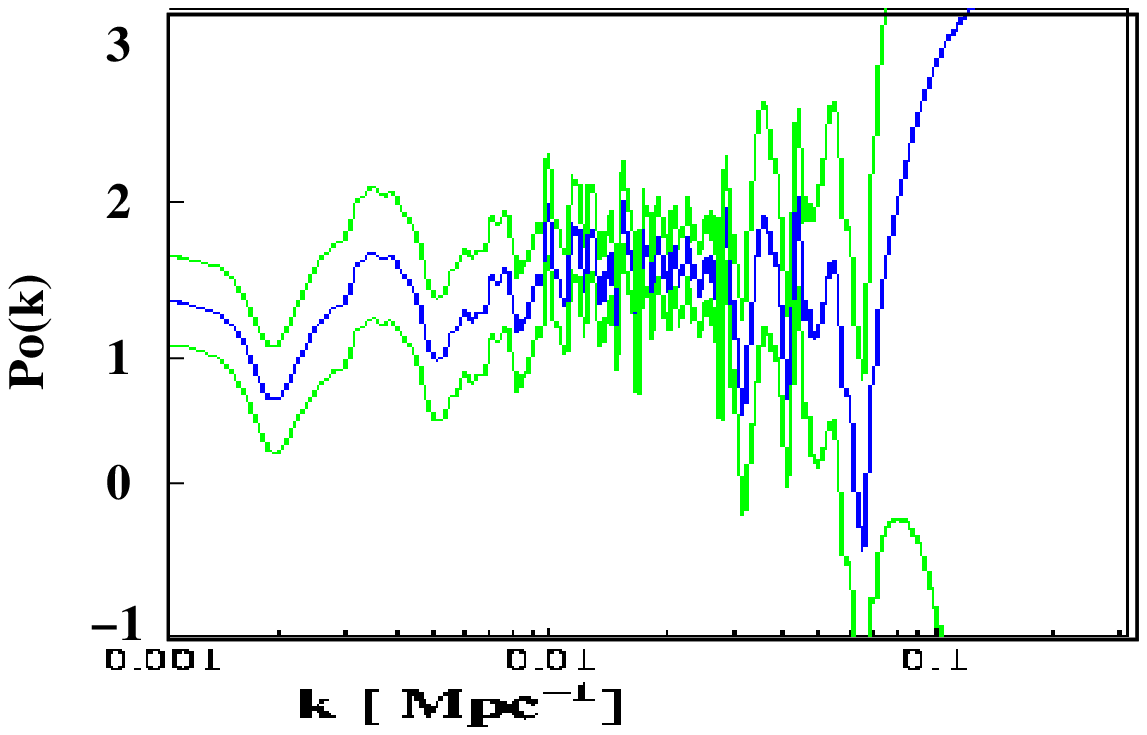}\hspace*{-8cm}\includegraphics[width=4cm, height=3.6cm]{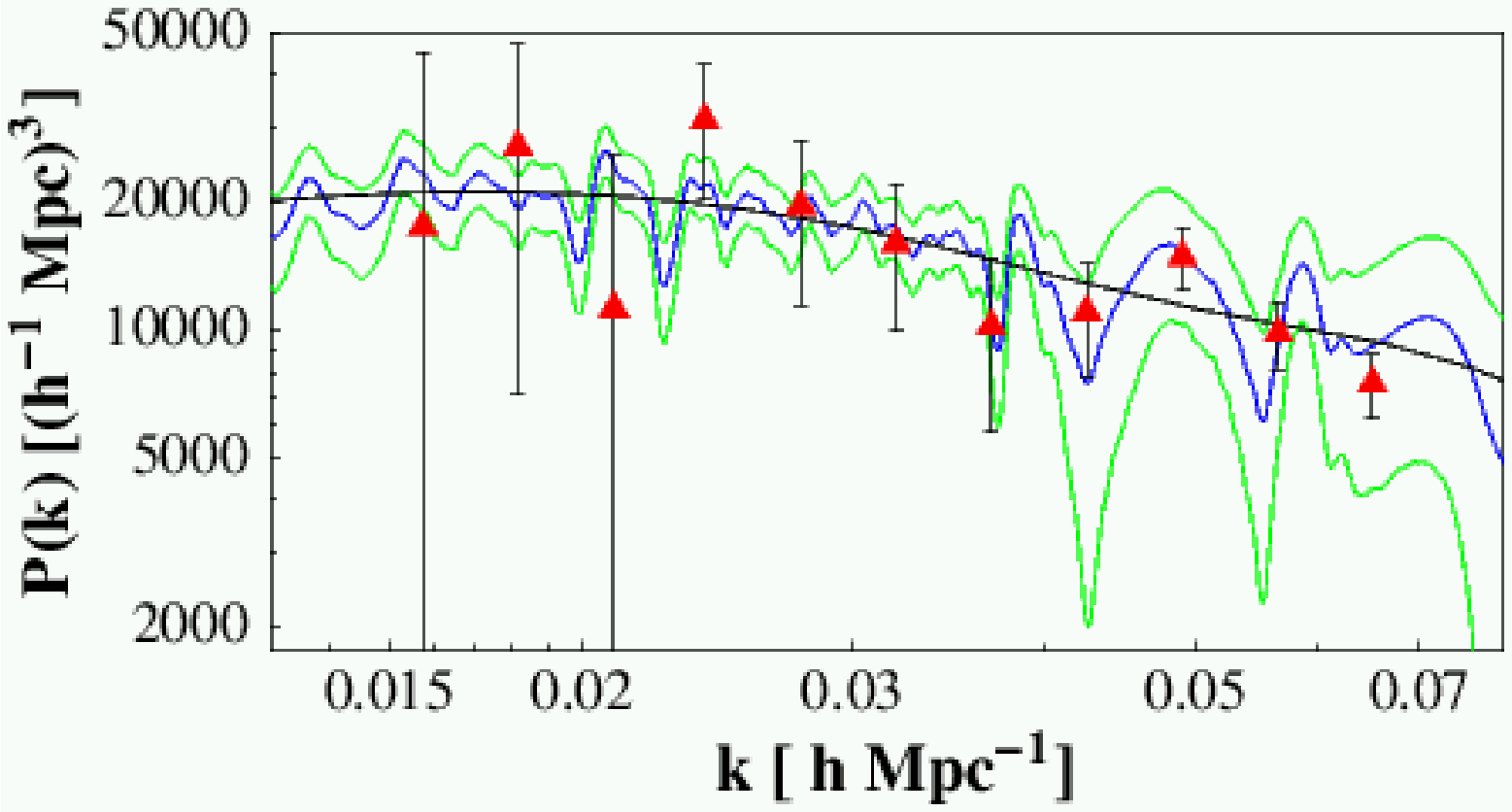}
\caption{Reconstructed IPS (left) from WMAP APS under concordance prior and corresponding matter power spectrum compared to SDSS data (middle). Example of reconstructed IPS from WMAP with $\Omega_{M}=1$ prior (right).}
\end{figure}

\subsection{From Combined Analysis}

In order to break these degeneracies, one can combine constraints from
different observations. Angular distances, shape of the matter power
spectrum, Hubble law are not sensitive to the cosmological parameters
in the same way as the APS of CMB anisotropies. This allows one to
break degeneracies intrinsic to one of the observable.  Figure 5
summarizes such a combination. Most of studies implying this kind of
combinations converge to the now standard ``concordant model''
(\cite{DRZB, tegmark,spergel}). Some observations remain nevertheless
inconsistent and combining them may lead to other conclusions
(\cite{hosz,BDRS,vauclair}). Such an example is given in the next
sections.

\begin{figure}[ht]
\centering
\includegraphics[width=7cm]{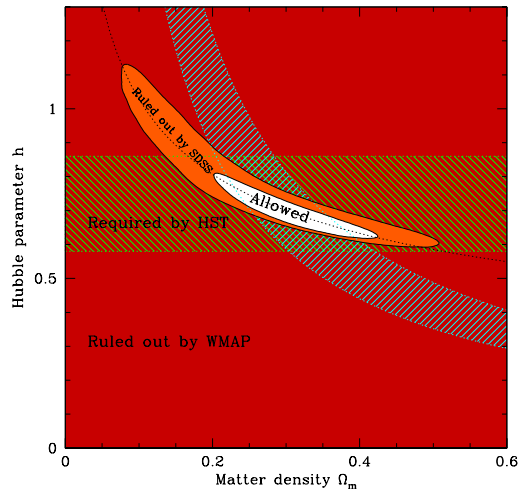}
\caption{Constraints ion the Hubble constant and matter density in combined analyses  (from Tegmark et al. 2003)}
\end{figure}

\section{Dark matter}

Dark matter is now a major component of the cosmological model. We
have direct proofs of existence (rotation curves of galaxy, mass
luminosity ratios, or lensing of background galaxy by clusters). This
dark matter is probably non baryonic (massive compact halo objects
search seems to indicate that such dark matter could only contribute
to 20\% of the total mass of clusters \cite{eros,macho}) and may
contribute to 25\% of the total density of the Universe (with 5\% of
baryons and 70\% of dark energy in the concordance model). 

This dark matter could be cold (few GeV, and non relativistic at
galaxy formation), warm (few keV) or hot (few eV and relativistic at
galaxy formation). Few candidates can be classified as follows
(\cite{gondolo}): existing (light neutrinos),well motivated (heavy
neutrinos, neutralinos, axions) and other (self-interactive
particles, wimpzillas, ...). Recent experiments have put a lower
limit on the mass of the neutrinos (\cite{klapdor}). On the other
side, heavy neutrinos have not been detected in laboratory, and the
shape of the matter power spectrum from CMB and galaxy surveys
(\cite{tegmark}) sets an upper bound to the mass of such particles. From
these limits one can deduce the cosmological contribution of neutrinos
in the energy density budget: $$ 0.0005 < \Omega_\nu h^2 < 0.03$$
whereas the total dark matter contribution is $\Omega_{DM}h^2\sim
0.12$. Neutrinos (light or heavy) can not therefore explain the full
dark matter content.  The next well motivated candidate is the
neutralino (cold dark matter). It is the lightest supersymetric
particle. It has not been detected yet but recent experiments (from
cosmology or particle physics) could not exclude it. See
\cite{gondolo} for a better an more complete review of non baryonic
dark matter and \cite{boehm} for alternatives to the supersymetric
candidates.

\section{The amplitude of fluctuations}

The amplitude of matter fluctuations in the present-day universe is an
important quantity of cosmological relevance. The abundance of
clusters is an efficient way to evaluate this quantity, commonly
expressed by $\sigma_8$, the {\em r.m.s.} amplitude of the matter
fluctuations on the $8 h^{-1}\rm Mpc$ scale. This quantity can also be
determined by the study of weak lensing shear or as a derived
parameter from CMB analysis.  The determination of $\sigma_8$ has
gained some tension because the amplitude of matter obtained from
clusters with hydrostatic equation leads to low values $\sigma_8 \sim
0.7 \pm 0.06$ (\cite{marke, reip, seljak})
while WMAP obtained recently $\sigma_8 \sim 0.9 \pm 0.1$ (\cite{spergel}). 

\subsection{From CMB}

WMAP team determination of cosmological parameter lead to the
concordance model. Such a model, normalized to CMB scales imply a value
of the amplitude of fluctuations $\sigma_8 \sim 0.9 \pm 0.1$. This
quantity is related to the amplitude of the APS in temperature around
$\ell \sim 1000$, whatever the shape of the IPS
(\cite{BDRS}). However, some degeneracies may alter this ``first
order'' assumption. Some physical processes which lower the power of
fluctuations at small scale may lower the value of $\sigma_8$. This is
the case of the reionisation of the Universe (or the presence of
neutrinos). Figure 7 shows the degeneracy between the optical depth of
reionisation ($\tau$) and the amplitude of fluctuations ($\sigma_8$)
when only the temperature APS is considered (which means no
constraints on $\tau$). For a Universe in a concordance model but
without reionisation, the amplitude of fluctuation can be as low as
$\sigma_8 \sim 0.65$. The cross power spectrum
(temperature--polarisation E) of WMAP contains a signature at large
scale of an early reionisation. This suggests a value of quite high
leading to the value of $\sigma_8$ quoted by \cite{spergel}. This
signature is nevertheless weak because some galactic foregrounds
contamination may play a role at these scales. The full polarized
spectrum (E--E), released soon, may clarify this.

\begin{figure}[ht]
\centering
\includegraphics[width=7cm]{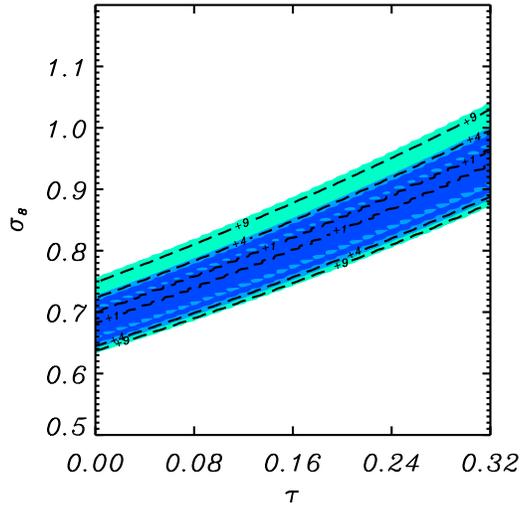}
\caption{Degeneracy between the amplitude of fluctuations and the optical depth from WMAP temperature APS.}
\end{figure}

\subsection{From weak lensing shear}

Weak lensing shear is another way to derive the amplitude of
fluctuations. Recent observations have lead to an ensemble of
consistent relations between the matter content and the amplitude of
fluctuations (\cite{BMRE, MLB02,Chang,Hamana,gems,HYG02,Jarvis,MRBE,
refre, RRG02, RhodesSTIS,VW}) which can be averaged by: $\sigma_8
\left( \frac{\Omega_M}{0.3} \right) ^{0.47} = 0.836\pm 0.035$.

\subsection{From Clusters of Galaxy}

A statistical precision
of few \% on $\sigma_8$ is possible from existing samples of x-ray
clusters, but in practice the relation between mass and temperature is
needed for such evaluation:
\begin{equation}
T =A_{TM} M^{2/3}_{15}(\Omega_M (1+\Delta_v)/179)^{1/3}h^{2/3}(1+z)\rm \;\; keV
\label{eq:tm}  
\end{equation}
This value of $A_{TM}$ has been estimated from x-ray properties of
clusters by different methods, essentially hydrostatic equation on one
side and numerical simulations on the other side, which lead to
sensitively different normalizations (from $\sigma_8 \sim 0.6$ to
$\sigma_8 \sim 1.$).

We have shown in \cite{BD04} a third way to derive the value of
$A_{TM}$ that we called self-consistent.

Clusters are useful cosmological probes in several important
ways. Their baryonic fraction $f_b$ can be inferred from observations:
$f_b = \frac{M_b}{M_{tot}}$.  Under the assumption that the baryonic
and dark matter amounts are representative of the universe, the baryon
fraction can be related to the cosmological parameters density
$\Omega_{b}$ and $\Omega_{m}$: $f_b = \Upsilon
\frac{\Omega_{b}}{\Omega_{m}}$ where $\Upsilon$ is a numerical factor
that has to be introduced in order to correct for the depletion of gas
during cluster formation and which can be determined only from
numerical simulations (\cite{white}). In practice a good working
value, at least in the outer part of clusters, is $\Upsilon = 0.925$
(\cite{frenk}).  The baryonic content of the Universe is now
known quite accurately through WMAP and other CMB measurements
($\omega_{b}= \Omega_{b}h^2 = 0.023\pm 0.002$, \cite{spergel})
the statistical uncertainty being doubled in order to account for
differences in various priors) essentially consistent with the
abundance of Deuterium (\cite{kirkman}) and with the baryonic
content of the IGM (\cite{tytler}).  We can furthermore assume
the following relation given by the position of the first Doppler peak
in the CMB APS (\cite{page}): $\Omega_m h^{3.4} =0.086\pm0.006$.  Then,
if we consider a sample of clusters around 4 keV from which we know
the baryon mass (\cite{vikhlinin}) we can rewrite Eq.~2 as: $A_{TM} =
f(\Omega_M)$, plotted in Figure 7 (red solid line with error bars
envelops). From there we conclude that $A_{TM}$ varies with $\Omega_M$
and is consistent with the baryon fraction in clusters whatever the
cosmology is, contrarily to the previous determinations (horizontal
shaded regions).

\begin{figure}[ht]
\centering
\includegraphics[width=7cm]{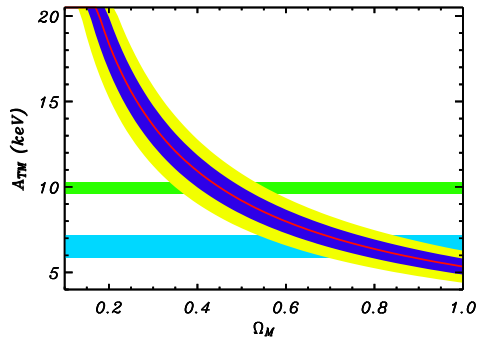}
\caption{The red line is the central value of 
$A_{\rm TM}$ which is the normalization of the mass temperature
relation Eq.~1. The WMAP relation between $H_0$ and
$\Omega_0$ has been used, as well as the constraint on the baryon
content of the Universe. One and two $\sigma$ uncertainties arising
from uncertainty on $\Omega_{b}$ ($= 0.023\pm 0.002$)  are shown as blue and yellow
areas. Horizontal areas correspond to estimations of $A_{\rm TM}$ from
hydrostatic methods (light green) obtained by \cite{RSB99} and 
\cite{marke} and from numerical simulations
(light blue) obtained from \cite{bryan, evrard}}
\end{figure}

Knowing the mass-temperature relation and its uncertainty we can
determine the amplitude of matter fluctuations by fitting the local
temperature distribution function and assuming a $\Gamma$-like
spectrum with $\Gamma = 0.2$. Here we use the Sheth and Tormen (1999)
mass function and a sample of x-ray selected local clusters ($f_x \leq
2.2 10^{-11}$ erg/s/cm$^2$ and $|b| \leq 20 \deg$, \cite{blanchard}
updated from BAX, \cite{bax}). The result is shown as lines in Figure
8.  As one can see, at a given value of $\Omega_M$ the amplitude of
$\sigma_8$ is well constrained.  Furthermore to the first order the
best $\sigma_8$ is independent of $\Omega_M$ ($\sigma_8 \sim 0.63 \pm
4.5\% $ for $\Omega_\Lambda = 0.7$.{ interestingly close to the value
obtained by \cite{viana}: $\sigma_8 \sim 0.61$ }). Our
conclusion appears somewhat surprising as it differs from standard
analyzes based on a fixed normalization $A_{TM}$, which cannot account
simultaneously for the baryon fraction in a consistent way for
arbitrary $\Omega_M$.

\begin{figure}[ht]
\centering
\includegraphics[width=7cm]{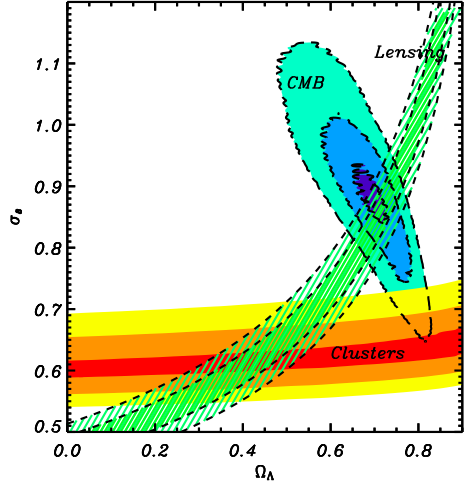}
\caption{The amplitude of matter fluctuations from clusters 
abundance using the mass--temperature relation found in the present
analysis compared to the amplitude of matter fluctuations derived from
CMB data (\cite{hinshaw, grainge, pearson, rhul, kuo}) grey area
correspond to 1,2,3 $\sigma$ contours on two parameters, dashed lines
are contours on one parameter. The one and two sigma amplitudes
obtained from an average of recent weak shear measurements are also
shown as dashed regions (see text for references).}
\end{figure}

\subsection{Evidence of New physics}

In the following we use the
constraint on $\sigma_8$ in a concordance model obtained from the CMB
fluctuations analysis including the temperature--polarization cross
power spectrum (TE) by the WMAP team.
The comparison of the value of $\sigma_8 $ from CMB data with the one
from clusters is revealing a critical discrepancy among the two
measurements (Figure 8).  It is clear that within any model with
$\Omega_\Lambda \sim 0.7$ the amplitude of $\sigma_8$ we derived from
clusters $\sigma_8 = 0.63 \pm 0.03$ is significantly smaller than what is 
expected from the CMB alone ($\sigma_8 = 0.88 \pm 0.035$).

As mentioned before, an accurate knowledge of $\tau$ is 
critical to properly evaluate the amplitude of matter fluctuations in
the concordance model.

Here above, we have considered models in which the dark matter is only
made of cold dark matter, the dark energy being a pure cosmological
constant (in term of the equation of state of vacuum $p = w \rho$,
this means $w = -1$), and that x-ray gas and known stars are the only
existing baryons in clusters. 

A first possibility to investigate is to examine whether a different
equation of state for the vacuum, so-called quintessence, might solve
this discrepancy.  We have therefore investigated flat models with
arbitrary $w$ and quintessence content $\Omega_Q$. Indeed combinations
of CMB and clusters data are known to provide tight constraints on
such models (\cite{DRZB}).  With the approach developed here,
models which were found to match CMB and clusters were found to
satisfy the following constraints: $0.46 <\Omega_Q < 0.54$ and $ - 0.5
< w < - 0.4 $.  Such models are currently at odds with constraints on
quintessential models (\cite{DRZB,tegmark,riess}) resulting from
combination of various data including the supernovae type Ia data. We
have therefore to turn toward more drastic paths to solve the above
issue.

In the following, we examine whether the introduction of an additional
component of the dark matter content of the universe would allow to
remove the above discrepancy. Neutrinos are known to exist and to be
massive, so perhaps the most natural massive component of the universe
to be introduced is in the form of a neutrino contribution.  This
solution has already been advocated in order to solve this discrepancy
in an Einstein de Sitter Universe (\cite{elgaroy, BDRS}). Indeed the
presence of a light, but non-zero, component of the dark matter
modifies significantly the transfer function of primordial
fluctuations which results in a lower amplitude on small scales. Given
existing measurements of mass differences we consider only the case
where the masses are equal. Within a concordance model
($\Omega_\Lambda = 0.7$, $\Omega_m = 0.3$) by combining the constraints
from CMB and clusters data, and marginalizing on ($\omega_{\rm b}$,
$H_0$, $n$, $\tau$) we found that a contribution of $\Omega_\nu =
0.015 \pm 0.01$ is preferred with a significance level, well above
3$\sigma$ (see Figure 9a), { improving the significance of such
possible evidence compared to Allen et al (2003)}.  This confirms that
the presence of a small contribution of neutrinos, with a typical mass
of .25 eV, to the density of the universe allows to reconcile the
amplitude of matter fluctuations from clusters with the one inferred
from CMB data. We notice that such value is above the upper limit
inferred by the WMAP team using a combination of several astronomical
data (\cite{spergel}). Finally, weak shear estimations have provided
measurements of the amplitude of matter fluctuations which not favor
such a solution.

 We are therefore left with the conclusion that our initial assumption
 that baryons in clusters are fairly representative of baryons in the
 universe is unlikely and therefore that the observed amount of baryon
 in clusters does not reflect the actual primordial value (a
 possibility that has been advocated by \cite{ettori}). Several
 mechanisms could lead to this situation: the most direct way could be
 the fact that a significant fraction of the baryons are in a dark
 form, either in the Universe or in clusters (for instance either in
 the form of Macho's, or in a large gaseous unidentified component,
\cite{bonamente}), or that a significant fraction of the baryons
 has been expelled from clusters during their formation process.  In
 such cases, the observed $M_b$ is biased low.  The actual mass of
 clusters  can then be obtained assuming a
 depletion factor $1-f$ implying that $f\Omega_b$ represents the
 missing baryons.  Again the combination of CMB and clusters
 constraints allow to evaluate the amplitude of $f\Omega_b$.  From
 Figure 9b, one can see such a component, $f\Omega_b \sim 0.023 $
 should represent nearly half ($\Upsilon \sim 0.5$) of the primordial
 baryon in order to solve the discrepancy. Although heating processes
 are advocated in order to account for observed properties of x-ray
 clusters, they currently do not lead to such a high level of depletion
 (\cite{bialek}).

\begin{figure}
\includegraphics[width=6cm]{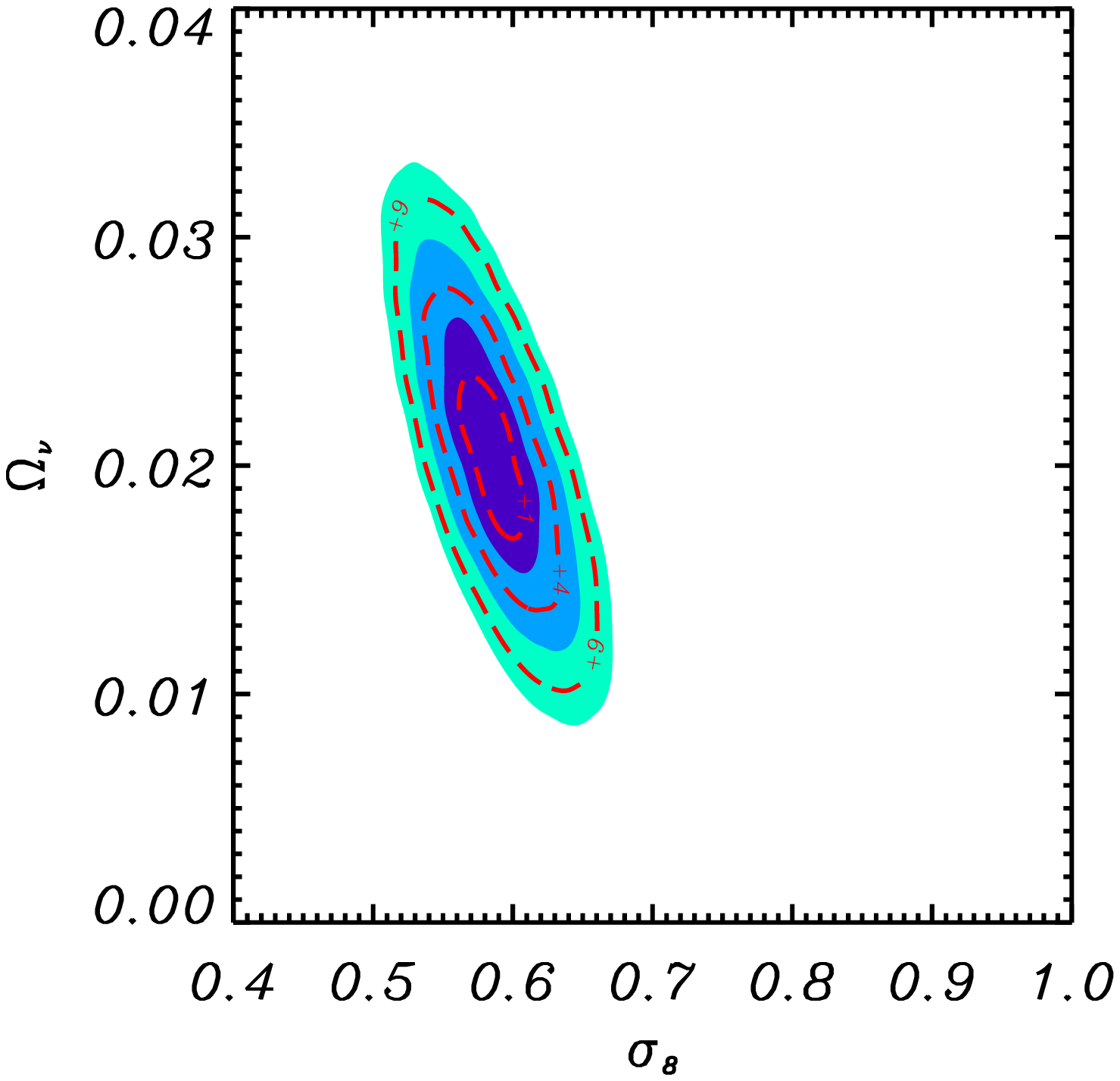}\includegraphics[width=6cm]{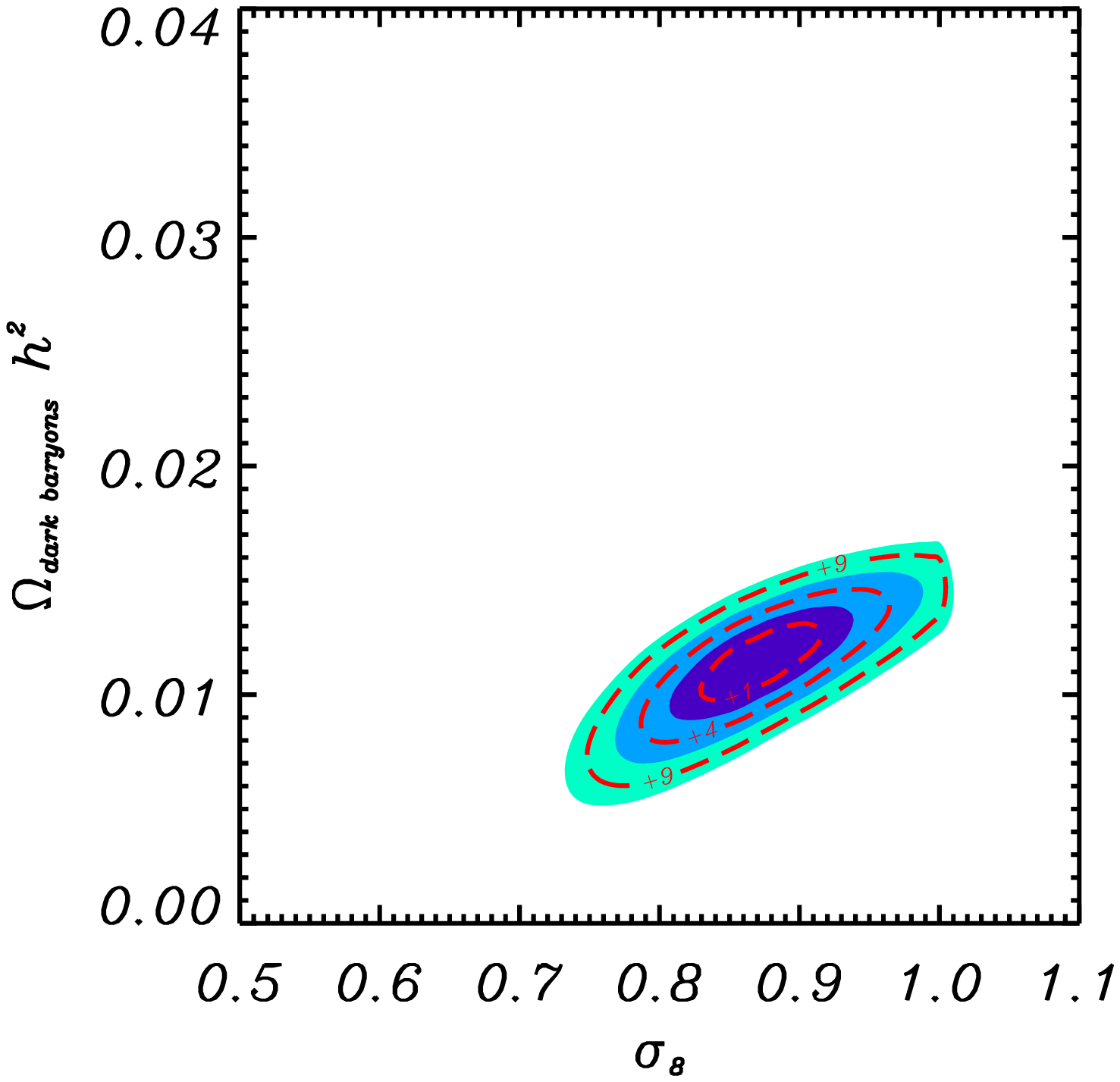}
\caption{Constraints on $\Omega_\nu$ and $\Omega_{dark\; baryon}h^2$ from a combined analyses of CMB and clusters, by fixing $\Omega_\Lambda=0.7$.
\label{powplot3}}
\end{figure}

\section{Summary}

We have seen how the concordance model seems to fit pretty well the
CMB power spectrum data, as well as a combination of other
cosmological observations. But it is also noticeable how much some
degeneracies may unable one to give strong constraints on cosmological
parameters. Furthermore, all the observations are not concordant.
The determination of the amplitude of matter fluctuations within pure
cold dark matter models, using two methods, namely the CMB and the
local clusters abundance, leads to two surprising significantly
different values.  There are several ways out to solve out this
discrepancy, although each of them represents a noticeable difference
with the standard concordance model.  The existence of a non-baryonic
dark component, like a neutrino contribution, would allow to solve
this discrepancy, although such a solution leads to a low value of
$\sigma_8$  which is not favored by some
other evidences.  If the actual value is actually larger, $\sigma_8
\sim 0.8-0.9$, the unavoidable conclusion is that the baryonic content
of clusters is not representative of the Universe.  In this case, an
astrophysical solution could be that baryons in clusters could be in a
dark form, or at least undetected until now. Alternatively, baryon in
clusters could have been severely depleted implying that the actual
value $\Upsilon$ is much smaller than the value we used above, the
apparent baryon fraction being biased low compared to the actual
primordial value.  Finally, several observations might help to clarify
this issue: the above conclusion relies on the actual value of the
optical depth $\tau$ found by WMAP. Other sources of informations on
$\sigma_8$ will also obviously put light on this issue: weak lensing
can potentially allow to measure directly the actual amplitude of
matter fluctuations with a similar precision to what has been obtained
here with clusters, provided that systematic uncertainties are fully
understood. Other direct measurements of the amplitude of matter
fluctuations like those derived from the Lyman--$\alpha$ forest power
spectrum (\cite{croft}) could also bring light on this issue. It
is remarkable that some of these observations that are expected in the
near future will potentially bring fundamental informations on
clusters physics { or alternatively} may reveal the existence of a
previously unidentified type of dark matter with $\Omega_{DM}$ as low
as 0.01.

\begin{acknowledgments}
 This research has made use of the X-Rays Clusters Database (BAX)
which is operated by the Laboratoire d'Astrophysique de Toulouse-Tarbes (LATT),
under contract with the Centre National d'Etudes Spatiales (CNES)
\end{acknowledgments}


\begin{chapthebibliography}{1}


\bibitem{eros} Afonso, C., et al.\ 2003, A\&A, 400, 951 

\bibitem{allen} Allen, 
S.~W., Schmidt, R.~W., \& Bridle, S.~L.\ 2003, MNRAS 346, 593 

\bibitem{BMRE}
Bacon D.,  Massey R.,  Refregier A.,    Ellis R.,  2003, MNRAS, 344, 673


\bibitem{bialek} Bialek, J.~J., 
Evrard, A.~E., \& Mohr, J.~J.\ 2001, ApJ, 555, 597 



\bibitem{BD04} Blanchard, A., \& Douspis, M.  2004, A\&A in press, astro-ph/0405489


\bibitem{BDRS} Blanchard, A., Douspis, M.,
Rowan-Robinson, M., \& Sarkar, S.\ 2003, A\&A 412, 35


\bibitem{blanchard} Blanchard, A., Sadat, R., Bartlett, J.~G., \& Le Dour, M.\ 2000, A\&A, 362, 809

\bibitem{boehm} Boehm, C., 
Fayet, P., \& Silk, J.\ 2004, PhrvD, 69, 101302 


\bibitem{bonamente} Bonamente, M., 
Joy, M.~K., \& Lieu, R.\ 2003, ApJ 585, 722 



\bibitem{MLB02}
Brown M.,  Taylor A.,  Bacon D.,  Gray M.,  Dye S.,  Meisenheimer K.,    Wolf
  C.,  2003, MNRAS, 341, 100


\bibitem{bryan}
Bryan, G.L. \& Norman, M.L. 1998, ApJ 495, 80

\bibitem{isoc} Bucher, M., Dunkley, J., 
Ferreira, P.~G., Moodley, K., \& Skordis, C.\ 2004, Physical Review 
Letters, 93, 081301 

\bibitem{Chang}
{Chang} T.-C.,  {Refregier} A.,  {Helfand} D.~J.,  {Becker} R.~H.,    {White}
  R.~L.,  2004, ApJ submitted, astroph/0408548


\bibitem{croft} Croft, R.A.C. Weinberg, D.H., Katz, N., Hernquist,  L.
\ 1998, ApJ 495, 44 

\bibitem{DRZB} Douspis, M., Riazuelo, A.,
Zolnierowski, Y., \& Blanchard, A.\ 2003, A\&A 405, 409

\bibitem{elgaroy} Elgar{\o}y,
{\O}.~\& Lahav, O.\ 2003, Journal of Cosmology and Astro-Particle Physics,
4, 4

\bibitem{ettori} Ettori, S. 2003, MNRAS 344, L13

 \bibitem{evrard}Evrard A.\ E., Metzler C.\ A. \& Navarro J.\ F., 1996, ApJ 469, 494 


\bibitem{fosalba}Fosalba, P.,  Szapudi, I., 2004, ApJ in press, astro-ph/0405589


\bibitem{frenk}Frenk, C.S, White, S.D.M et al. 1999, ApJ 525, 554


\bibitem{gondolo}Gondolo, P., 2004, astro-ph/0403064

\bibitem{gorski} Gorski, K.~M., Hinshaw, 
G., Banday, A.~J., Bennett, C.~L., Wright, E.~L., Kogut, A., Smoot, G.~F., 
\& Lubin, P.\ 1994, ApJL, 430, L89 

\bibitem{grainge} Grainge, K., et al.\ 
2003, MNRAS 341, L23 


\bibitem{Hamana}
{Hamana} T.,  {Miyazaki} S.,  {Shimasaku} K.,  {Furusawa} H.,  {Doi} M.,
  {Hamabe} M.,  {Imi} K.,  {Kimura} M.,  {Komiyama} Y.,  {Nakata} F.,  {Okada}
  N.,  {Okamura} S.,  {Ouchi} M.,  {Sekiguchi} M.,  {Yagi} M.,    {Yasuda} N.,
  2003, ApJ, 597, 98




\bibitem{gems} {Heymans} C. et al., 2004,  astroph/0406468

\bibitem{hinshaw} Hinshaw, G., et al.\ 
2003, ApJ supp., 148, 135 



\bibitem{HYG02}
Hoekstra H.,  Yee H.,    Gladders M.,  2002, ApJ, 577, 595



\bibitem{Jarvis}
Jarvis M.,  Bernstein G.,  Jain B.,  Fischer P.,  Smith D.,  Tyson J.,
  Wittman D.,  2003, ApJ, 125, 1014


\bibitem{kirkman} Kirkman, D., Tytler,
D., Suzuki, N., O'Meara, J.~M., \& Lubin, D.\ 2003, ApJS 149, 1

\bibitem{klapdor} Klapdor-Kleingrothaus, H.~V., 
Krivosheina, I.~V., Dietz, A., \& Chkvorets, O.\ 2004, Physics Letters B, 
586, 198 


\bibitem{hosz} Kochanek, 
C.~S.~\& Schechter, P.~L.\ 2004, Measuring and Modeling the Universe, 117 


\bibitem{kuo} Kuo, C.~L., et al.\ 2004, 
ApJ 600, 32 

\bibitem{lineweaver} Lineweaver, C.~H., Barbosa, D., 
Blanchard, A., \& Bartlett, J.~G.\ 1997, A\&A, 322, 365 

\bibitem{MRBE} Massey, R.,  Refregier, A. Bacon, D., Ellis, R. \  2004, astroph/0405457

\bibitem{marke} Markevitch, A. \ 1998, ApJ   504, 27


\bibitem{navarro} Navarro, J.\ F., Frenk, C.\ S. \& White, S.\ D.\ M. 1995, MNRAS 275,
720 (NFW)


\bibitem{page} Page, L., et al.\ 2003,
ApJS 148, 233

\bibitem{pearson} Pearson, T.~J., et al.\ 
2003, ApJ 591, 556 

\bibitem{refre} Refregier, A.\ 2003, ARA\&A 
41, 645 



\bibitem{RRG02}
{Refregier} A.,  {Rhodes} J.,    {Groth} E.~J.,  2002, ApJL, 572, L131



\bibitem{reip} Reiprich, T.~H., \& B\"ohringer, H.\  2002, ApJ 567, 716

\bibitem{riess} Riess, A.G. et al.\ 2004, ApJ  607, 665 


\bibitem{RhodesSTIS}
{Rhodes} J.,  {Refregier} A.,  {Collins} N.~R.,  {Gardner} J.~P.,  {Groth}
  E.~J.,    {Hill} R.~S.,  2004, ApJ, 605, 29

\bibitem{RSB99} Roussel, H., Sadat, R. \& Blanchard, A. 2000, A\&A 361, 429

\bibitem{rhul} Ruhl, J.~E., et al.\ 2003,  ApJ 599, 786 

\bibitem{sadat2} Sadat, R., \& Blanchard, A.\ 2001, A\&A 371, 19



\bibitem{bax} Sadat, R., Blanchard, A., Kneib, J.-P., Mathez, G., Madore, B., \& Mazzarella, J.~M.\ 2004, A\&A, 424, 1097



\bibitem{sdss} www.sdss.org


\bibitem{seljak} Seljak, U.\ 2002, MNRAS 337, 769


\bibitem{ST} Sheth, R.~K. \& Tormen, G.\ 1999, MNRAS 308, 119


\bibitem{spergel} Spergel, D.~N.~et al.\
2003, ApJS, 148, 175

\bibitem{tegmark} Tegmark, M. et al., 2004, astro-ph/0310723, Phys. Rev. D  69, 103501 



\bibitem{TDS} Tocchini--Valentini, D., Douspis, M., \& Silk, J., 2004, MNRAS in press, astro-ph/0402583



\bibitem{tytler} Tytler, D. et al.,  2004, astro-ph/0403688 

\bibitem{vikhlinin}
{Vikhlinin}, A., {Forman}, W. and {Jones}, C. 1999, ApJ 525, 47

\bibitem{VW} Van Waerbeke, L., 
Mellier, Y., Hoekstra, H. \ 2004, A\&A, 393, 369 

\bibitem{vauclair} Vauclair, S.~C., et al.\ 2003, A\&A, 412, L37 


\bibitem{viana} Viana, 
P.~T.~P., Nichol, R.~C., \& Liddle, A.~R.\ 2002, ApJL 569, L75 

\bibitem{white}
White, S.D.M., Navarro, J.F., Evrard, A.E. \& Frenk, C. 1993, Nature 366, 429

\bibitem{wmap} lambda.gsfc.nasa.gov/product/map

\bibitem{macho} Yoo, J., 
Chanam{\' e}, J., \& Gould, A.\ 2004, ApJ, 601, 311 

\end{chapthebibliography}
\end{document}